\documentstyle[12pt,epsf]{article}
\newcommand{\be}{\begin{equation}}
\newcommand{\ee}{\end{equation}}
\newcommand{\al}{\mbox{$\alpha$}}

\newcommand{\s}{\mbox{$\sigma$}}

\newcommand{\bi}[1]{\bibitem{#1}}
\newcommand{\fr}[2]{\frac{#1}{#2}}

\newcommand{\gm}{\mbox{$\gamma_{\mu}$}}

\newcommand{\GD}{\mbox{$\tilde{G}$}}
\newcommand{\gf}{\mbox{$\gamma_{5}$}}

\begin{document}
\begin{titlepage}
\rightline{\vbox{\halign{&#\hfil\cr
&UQAM-PHE-97/04\cr
&\today\cr}}}
\vspace{0.5in}
\begin{center}

\Large\bf  CP-odd interaction of axion with matter
\\
\medskip
\vskip0.5in

\normalsize  
{\bf M. Pospelov}\footnote{pospelov@mercure.phy.uqam.ca} 

\smallskip
\medskip

{\em D\'{e}partement de Physique, Universit\'{e} du Qu\'{e}bec a 
Montr\'{e}al\\ 
C.P. 8888, Succ. Centre-Ville, Montr\'{e}al, Qu\'{e}bec, 
Canada, H3C 3P8}\\
and\\
{\em Budker Institute of Nuclear Physics, Novosibirsk, 
630090, Russia }
\smallskip
\end{center}
\vskip1.0in

\noindent{\large\bf Abstract}
\smallskip\newline
In the world with axion the precise measurement of gravitational force and check of the equivalence principle at small distances, $\lambda\sim 1cm$ and less, could provide an additional test of CP symmetry. Using the chiral approach, we  connect the strength of CP-odd axion-nucleon vertex with CP-odd pion-nucleon coupling for color electric dipole moments of quarks as a low energy source of CP violation. Strong limits on CP-odd coupling $g_{\pi NN}$ coming from atomic electric dipole moments show that gravitational experiments are six orders of magnitude away from the sensitivity level needed to detect axion-mediated forces. 
\end{titlepage}

\section{Introduction}

The strong CP problem is one of the most intriguing issues of modern particle 
physics. The additional term in the QCD Lagrangian 
\be
{\cal L}= \theta\fr{g^2}{32\pi^2} G^a_{\mu\nu}\GD^a_{\mu\nu} 
\ee
violates P and CP symmetries \cite{theta}. In the electroweak theory, a 
diagonalization of quark mass matrices $M_u$ and $M_d$ involves chiral 
rotations and brings the additional contribution to  theta term:
\be
\bar{\theta}=\theta+arg(det M_uM_d)
\ee

Current experimental limits on the electric dipole moment (EDM) of the 
neutron \cite{nEDM} put a severe constraint on  $\bar{\theta}$ parameter. The chiral 
algebra calculation of the neutron EDM induced by the theta term  \cite{CDVW} 
gives the following prediction:
\be
d_n\simeq 3.6\times10^{-16}\bar{\theta}\,e\cdot cm.
\label{eq:dnt}
\ee
Together with the current neutron EDM constraints it implies the limit 
$\bar{\theta}<3\cdot10^{-10}$.

The puzzling smallness of  $\bar{\theta}$ in comparison with a naive expectation $\theta\sim 1$ is usually referred to as strong CP problem. It could be solved theoretically in 
different manners. One can postulate that the starting point for $\theta$-parameter is zero and mass matrices are hermitean at some high-energy scale due to some symmetry reasons \cite{Moh}. The evaluation down to the electroweak scale must keep radiative corrections to $\bar\theta$ under the level of $10^{-9}$. This scenario can hardly accommodate any sizable CP-violating phase beyond CP violation coming from Yukawa matrices (except, maybe, lepton sector). In all known models of CP violation which predict large EDMs and color EDMs (CEDMs) of quarks the radiative corrections to $\bar\theta$ give the EDM of the neutron already larger than current experimental bound. In the Minimal Supersymmetric Standard Model (MSSM), for example, this implies that a hidden dynamics responsible for supersymmetry breaking must respect CP to produce real soft-breaking parameters \cite{DGH}. 

In contrast, the existence of axion mechanism allows the dynamical relaxation of $\bar{\theta}$ \cite{PQ,DSZ} {\em and} leaves a room for large CP violating effects induced by the effective operators dim$\geq 5$. 

The combination of CP-violation and PQ symmetry leads to a very interesting phenomenon, the long range interaction mediated by axion \cite{MW}. It is shown schematically at Fig. 1. The axion mass of $10^{-5}$ eV corresponds to the characteristic length for this interaction of about 2 cm. It mimics gravity at these distances and therefore could be traceable in the measurement of gravitational forces at small distances \cite{Hosk}. Moreover, this new interaction needs not respect equivalence principle and can show up in the  precise E\"otv\"os-Braginsky type of experiment in sub-cm region \cite{Adel}. The latter is very promising because accuracy of this check at large distances can be as good as $10^{-12}$.

In Standard Model, though, these effects are known to be negligibly small \cite{MW,G}. In different modifications of supersymmetric models with new CP-violating sources besides the conventional Kobayashi-Maskawa phase there is a certain hope that the new long-range interaction can be detected by new precision measurements of gravitational forces at the small distances \cite{DT},\cite{BRS}. Using the Naive Dimensional Analysis (NDA), the authors of work \cite{BRS} connected the strength of axion mediated force with maximal CP violation allowed by the limit on the electric dipole moment of neutron. Their result suffers from a rather large uncertainty (four orders of magnitude) and do not include the analysis of possible equivalence principle violation. In this letter we calculate the strength of the axion-matter coupling and dynamically induced theta term using chiral perturbation theory and some elements of the QCD sum rules. Our aim is to improve the accuracy of the predictions based on NDA and connect this vertex with CP-odd pion-nucleon coupling constant, strongly limited from the experiments searching for EDMs of diamagnetic atoms \cite{EDM} (For the detailed analysis of this coupling, see Refs. \cite{KKY,KK}.) 

\section{CP-odd interaction of the axion with matter}

Motivated by different supersymmetric models, we choose CEDMs as the most important effective operators and neglect possible effects from Weinberg  three gluon operator and four-fermion type of interaction. 
Then the relevant part of the effective Lagrangian at the scale of 1 GeV is
\be
{\cal L}_{eff}= +\fr{a}{f_a}\fr{\al_s}{8\pi}G^a_{\mu\nu}\GD^a_{\mu\nu}+ i\sum_{i=u,d,s}\fr{\tilde{d}_i}{2}\bar{q}_igt^aG^a
_{\mu\nu}\s_{\mu\nu}\gf q_i
+...~,
\label{eq:eff}
\ee
were $a$ is the axion field and $f_a$ is the axion decay constant. Dots stand here for the model dependent gradient interaction, $\partial_\mu a\cdot j_\mu$, which can generate only contact interaction and therefore is irrelevant for our problem.  

In the absence of CP violation, nonremovable by PQ transformation, PQ symmetry sets theta parameter to zero \cite{PQ}. The situation is different in the presence of extra CP-violating sources, communicated by the operators dim$\ge5$. Thus, CEDMs of quarks drive theta parameter from zero to a value given by a ratio of two correlators:
\begin{eqnarray}
\theta_{eff}=-\fr{K_1}{|K|}, \;\;\mbox{where}\;\;
K=i\left\{\int dx e^{ikx}\langle 0|T(\fr{\al_s}{8\pi}G\GD(x),
\fr{\al_s}{8\pi}G\GD(0))|0 \rangle\right\}_{k=0}\\
K_1=i\left\{\int dx e^{ikx}\langle 0|T(\fr{\al_s}{8\pi}G\GD(x),
i\sum_{i}\fr{\tilde{d}_i}{2}\bar{q}_ig(G\s)\gf q_i(0)
|0 \rangle\right\}_{k=0}\nonumber
\end{eqnarray}
and $G^a
_{\mu\nu}\GD^a
_{\mu\nu}\equiv G\GD$, $(G\s) \equiv t^aG^a
_{\mu\nu}\s_{\mu\nu}$. The nonzero value of $\theta_{eff}$, in its turn, induces CP-violating axion-nucleon coupling \cite{MW}:
\be
g_{aNN}=\fr{\theta_{eff}}{f_a}\langle N|\fr{m_dm_sm_u}{m_sm_d+m_sm_u+m_dm_u}(\bar uu +\bar dd +\bar ss)|N\rangle.
\ee
Even though we are already able to set a constraint on this vertex using (\ref{eq:dnt}), it is still instructive to calculate $K_1$ and $\theta_{eff}$.

The calculation of $K$ is based on the use of the anomaly equation and the saturation of subsequent quark pseudoscalar correlator by light hadronic states \cite{SVZ}. It yields a formula for the axion mass:
\be
m_a^2=-\fr{K}{f_a^2}=\fr{m_\pi^2f_\pi^2}{f_a^2}
~\fr{m_um_d}{(m_u+m_d)^2}.
\ee
Same technique can be exploited in the case of $K_1$ (for the case of $m_u=m_d$ the explicit derivation of $K_1$ can be found in Ref. \cite{BU}). Using the anomaly equation in the form proposed in \cite{BT},
\begin{eqnarray}
\partial_\mu\fr{m_dm_s\bar u\gamma_\mu\gamma_5u +m_um_s
\bar d\gamma_\mu\gamma_5d+m_um_d\bar s\gamma_\mu\gamma_5s}{m_sm_d+m_sm_u+m_dm_u}=\nonumber\\
\fr{2m_um_dm_s}{m_sm_d+m_sm_u+m_dm_u}
(\bar u\gamma_5u+\bar d\gamma_5d+\bar s\gamma_5s)+\fr{\al_s}{4\pi}G\GD,
\label{eq:anom}
\end{eqnarray}
we apply standard technique of current algebra. The correlator of interest, $K_1$, can be rewritten in the form of the equal-time commutator, which we calculate easily, plus a term containing the singlet combination of pseudoscalars built from quark fields: 
\begin{eqnarray}
K_1=-\fr{1}{2}\langle 0|\fr{m_dm_sm_u}{m_sm_d+m_sm_u+m_dm_u}
\left(\fr{\tilde d_u}{m_u}\bar u g(G\s)u+
\fr{\tilde d_d}{m_d}\bar d g(G\s)d+\fr{\tilde d_s}{m_s}\bar s g(G\s)s\right)|0 \rangle+\\\nonumber
\int d^4x\langle 0|
T\{\fr{im_um_dm_s}{m_sm_d+m_sm_u+m_dm_u}
(\bar u\gamma_5u+\bar d\gamma_5d+\bar s\gamma_5s)(x),
i\!\sum_{i}\fr{\tilde{d}_i}{2}\bar{q}_ig(G\s)\gf q_i(0)\}|0 \rangle
\end{eqnarray}
The second term here is suppressed by an extra power of light quark masses in the numerator. It would bring a comparable contribution, though, if there were an intermediate hadronic state with a mass, vanishing in the chiral limit $m_i\rightarrow 0$. At the same time the flavor structure of this term shows that the lightest intermediate state here is $\eta'$ which is believed to remain heavy even if quark masses vanish. Thus the contribution from the second term is negligible in the limit $m_{\pi}\ll m_{\eta'}$. 

Finally we get the dynamically induced theta term in the following compact form:
\be
\theta_{eff}=-\fr{m_0^2}{2}\left(\fr{\tilde d_u}{m_u}+
\fr{\tilde d_d}{m_d}+\fr{\tilde d_s}{m_s}\right).
\ee
$m_0^2$ here is the ratio of the quark-gluon condensate to the quark condensate. It is known to sufficient accuracy from the baryon sum rules \cite{SR},
\be
m_0^2=\fr{\langle 0|g \bar q (G\sigma)q|0\rangle}
{\langle 0| \bar q q|0\rangle}\simeq 0.8\mbox{GeV}^2.
\ee

Now we turn to the calculation of the axion-nucleon CP-violating coupling not related to $\bar{\theta}$. This coupling is given by the matrix element over the nucleon state of the structure similar to $K_1$:
\be
i\int dx\langle N| \left\{e^{ikx}T(\fr{\al_s}{8\pi}G\GD(x),
i\sum_{i=u,d,s}\fr{\tilde{d}_i}{2}\bar{q}_it^a(G^a\s)\gf q_i(0)
\right\}_{k=0}|N \rangle
\label{eq:start}
\ee 
Using the equation (\ref{eq:anom}), we go along the same line as in the calculation of $\theta_{eff}$ and connect Eq. (\ref{eq:start}) with the matrix elements of the dipole-type operators over nucleon. The new thing here is the existence of additional contributions related with the diagrams of Fig. 2. Taking into account all contributions, after some algebra, we arrive at the following formula:
\begin{eqnarray}
\fr{m_dm_sm_u}{m_sm_d+m_sm_u+m_dm_u}
\left[-\fr{1}{2}\langle N|\left(\fr{\tilde d_u}{m_u}\bar u(G\s) u+
\fr{\tilde d_d}{m_d}\bar d(G\s) d+\fr{\tilde d_s}{m_s}\bar s(G\s) s\right)|N\rangle\nonumber+\right.\\\left.
m_0^2\langle N|
\fr{1}{2}
(\bar uu-\bar dd)\fr{\tilde d_u-\tilde d_d}{m_u+m_d}+
(\bar uu+\bar dd -2\bar ss)\fr{\tilde d_u+\tilde d_d-2
\tilde d_s}{m_u+m_d+4m_s}|N\rangle\right]
\label{eq:master}
\end{eqnarray}
It should be noted here that the CP-odd nature of the interaction $a\bar NN$ was crucial for the transformation of Eq. (\ref{eq:start}) to the series of matrix elements from local operators (\ref{eq:master}). It allows to drop all "nonlocal" terms proportional to $k_\mu\langle N|T(\bar q \gm\gf q(x),i\bar q g(G\s)\gf q(0)|N\rangle$ as leading to $k^2$ in the numerator. It would be not the case, for example, if we were calculating corrections to CP-even interaction $ia\bar N\gf N$ caused by $\bar q g(G\s) q$ where similar matrix elements are not negligible. 

Further progress is related with the application of low energy theorem \cite{VZS} in $0^+$ channel to the matrix element of the quark-gluon operator
\be
\langle N|\bar q g(G\sigma)q|N\rangle\simeq \fr{5}{3}m_0^2\langle N|\bar qq|N\rangle.
\label{eq:let}
\ee
For $\langle N|\bar qq|N\rangle$ we have the following numbers for certain flavors \cite{Z}:
\be
\langle p|\bar uu|p\rangle\simeq 4.8;\;\;\langle p|\bar dd
|p\rangle\simeq 4.1;\;\; \langle p|\bar ss|p\rangle\simeq 2.8
\label{eq:me}
\ee
with the obvious generalization on the neutron. 

From these relations we are able to learn one unfortunate thing. The dynamics of strong interaction makes CP-violating coupling of axion with neutron and proton be nearly equal even if CP violation at small distances is isospin asymmetric (exept, maybe, some very specific cases),
\be
g_{app}-g_{ann}\ll g_{app}+g_{ann}\equiv 2g_{aNN}
\label{eq:p-n}
\ee
The strength of isospin-symmetric coupling $g_{aNN}$ follows from Eq. (\ref{eq:master}) after the substitution of (\ref{eq:let})-(\ref{eq:me}),
\be
g_{aNN}\simeq \fr{1}{f_a}\fr{m_um_d}{m_u+m_d}m_0^2\left(3.8(\fr{\tilde d_u}{m_u}+\fr{\tilde d_d}{m_d})-0.7\fr{\tilde d_s}{m_s}\right)
\label{eq:answer}
\ee

At this point we should plug in the limits on $\tilde d_i$ coming from the EDM-type of experiments. As it was shown in \cite{KKY,KK}, the data on atomic EDMs set better limits on $\tilde d_u$ and $\tilde d_d$ than neutron EDM does. For us, however, it is more important that $g_{aNN}$ can be connected with pion-nucleon CP-odd vertex with little degree of uncertainty. This vertex, $g_{\pi NN}$, leads to the T-odd nucleon-nucleon interaction $\bar N N\bar N'i\gf N'$, to Schiff moment of nuclei and ultimately to an atomic EDM. This calculation was done in Refs. \cite{KKY,KK} by means of the same technique that we exploit here and we simply quote this result:
\be
g_{\pi NN}=\fr{1}{2f_\pi}\langle N|\tilde d_u\bar ug(G\s) u - \tilde d_d\bar dg(G\s) d |N\rangle \simeq \fr{1}{f_\pi}\fr{5}{6} m_0^2(\tilde d_u-\tilde d_d)\langle N|\fr{\bar uu +\bar dd}{2} |N\rangle.
\label{eq:pion}
\ee
Again, $\langle N|\bar uu -\bar dd |N\rangle$ was neglected in comparison with $\langle N|\bar uu +\bar dd |N\rangle$.
Formulae (\ref{eq:answer})-(\ref{eq:pion}) solve, in principle, the problem of correspondence between $g_{\pi NN}$ and $g_{aNN}$. However, instead of writing a general formula connecting these two quantities, it is convenient to go to some specific and more instructive physical cases making some assumptions about underlying CP-violating physics. 

{\em Case 1. Proportionality}\newline
\mbox{} By "proportionality" we understand a certain situation when $\tilde d_u/m_u=\tilde d_d/m_d=\tilde d_s/m_s$. It happens, for example, in MSSM when the relative CP-violating phase between trilinear soft-breaking parameter and gaugino mass is mediated by gluino exchange diagram. 

In this case the relation between two couplings is given by the following formula:
\be
g_{aNN}=g_{\pi NN}\fr{f_\pi}{f_a}\fr{2m_um_d}{m_d^2-m_u^2}
\label{eq:prop}
\ee
As to the potential violation of the equivalence principle by this interaction, it is given by the following relation:
\be
\fr{g_{ann}-g_{app}}{g_{ann}+g_{app}}=\fr{\langle p|\bar uu -\bar dd|p\rangle}{\langle p|\bar uu +\bar dd|p\rangle}\simeq 0.08.
\ee

{\em Case 2. One-flavor dominance}\newline
\mbox{} This is an opposite to the previous case situation when $\tilde d_i/m_i$ is enhanced for certain flavor. This may happen within different Grand Unification scenarios \cite{DH}.

For u(d)-dominance the connection between $g_{aNN}$ and $g_{\pi NN}$ is unambiguous; it follows from general formulae (\ref{eq:master}) and (\ref{eq:pion}) {\em before} taking matrix elements, so that major QCD-related uncertainties are gone. The resulting relation takes the simplest possible form:
\be
g_{aNN}=g_{\pi NN}\fr{f_\pi}{f_a}\fr{m_{d(u)}}{m_u+m_d},
\label{eq:flav}
\ee
where the subscripts $d(u)$ correspond to the case of up(down)-flavor dominance. 

If $\tilde d_u-\tilde d_d$ is close to zero (although, we do not know explicit examples of such a model), the coupling $g_{\pi NN}$ becomes small. It does not mean, of course, that $g_{aNN}$ can grow uncontrollably on account of the limit from neutron EDM which is sensitive to $\tilde d_u+\tilde d_d$ combination \cite{KK}.

\section{Comparison with experimental data}
The experimental bound on EDM of $^{199}$Hg \cite{EDM} translates into the following stringent limit on $g_{\pi NN}$
\be
g_{\pi pp}\simeq g_{\pi NN}< 2\cdot 10^{-11}
\ee
In the most interesting case of proportionality we get the following bound on the ratio of axion-mediated long-range interaction to gravitational force for the range of new force $\lambda=m_a^{-1}$:
\be
\fr{F_{axion}}{F_{gravity}}=\fr{g_{aNN}^2}{4\pi \gamma m_N^2}=5\cdot 10^{11}g_{\pi NN}^2\left(\fr{1cm}{\lambda}\right)^2< 2\cdot 10^{-10}\left(\fr{1cm}{\lambda}\right)^2.
\label{eq:ratio}
\label{eq:3a}
\ee
This ratio is tremendously small as compared to the current experimental sensitivity at the level of $10^{-4}$. 

More promising in terms of experimental accuracy is a high-precision check of the equivalence principle at small distances. The existing experiment sets the limit on the relative accelerations of Cu and Pb toward a uranium attractor as precise as $10^{-8}$ at the distances $\lambda \ge 10~cm$ \cite{Adel}. What can we expect from the interaction mediated by axion in this case? For the case of maximal isospin-asymmetric CP-violation ($\tilde d_u$-dominance) the estimate of the equivalence principle violation by axion mediated forces takes the following form:
\be
\fr{a_{Cu}-a_{Pb}}{a_{gravity}}\simeq \fr{(N-Z)_{Pb}}{2A_{Pb}}~\fr{\langle p|\bar uu- \bar dd|p\rangle}{\langle p|\bar uu+ \bar dd|p\rangle} ~\fr{F_{axion}}{F_{gravity}}<2\cdot 10^{-12}\left(\fr{1cm}{\lambda}\right)^2.
\label{eq:3b}
\ee
The results (\ref{eq:3a})-(\ref{eq:3b}) are summarized in Fig 3.

\section{Conclusions} 
In the case of CP violation communicated by color EDMs, the axion-nucleon CP-violating vertex is calculated by means of the current algebra. 
We have shown that the sensitivity of current gravitational experiments at small, $\sim$ 1cm, distances needs to be improved  by six orders of magnitude to probe the CP-violating interaction of axion with matter.
Our conclusion is very firm and do not contain a large degree of uncertainty reported in the previous analysis \cite{BRS}. It is based on the relations between $g_{aNN}$ and $g_{\pi NN}$, Eqs. (\ref{eq:prop})-(\ref{eq:flav}), which are held to better accuracy than the calculation of $m_a$, $g_{aNN}$ and $g_{\pi NN}$ itself. Although, color EDMs are, perhaps, the most interesting CP-violating operators, one can implement similar procedure for all variety of CP-odd operators dim=6 built from the quark fields. For Weinberg operator as a source, though, we are not able to give a better estimate than that based on NDA.

Our result for $F_{axion}/F_{gravity}$, Eq. (\ref{eq:3a}), is significantly lower than the central value of the prediction based on NDA and neutron EDM data \cite{BRS}. Once again it shows usefulness of the limits on CP-violation extracted from  EDM of $^{199}$Hg \cite{EDM}.

Author would like to thank W. Marciano, V. Miransky, R. Pisarsky and M. Tytgat for helpful discussions and Theory group of BNL for the hospitality extended to him during his visit where the part of this work was done. This work is supported by NATO Science Fellowship, N.S.E.R.C., and Russian Foundation for Basic Research.

\newpage
{\bf Figure captions.}

Fig. 1 The symbolic picture of axion exchange between two nucleons with CP violation in both vertices.

Fig. 2 Additional contribution to $g_{aNN}$, not related to $\theta$ or $\langle N|\bar q g(G\sigma)q|N\rangle$.

Fig. 3a The comparison of maximal strength for $F_{axion}/F_{gravity}$ (straight line) with current experimental sensitivity (curve in the upper corner) \cite{Hosk}.

Fig. 3b The comparison of predicted maximal strength for $\Delta a/a$ with the limits from \cite{Adel}.
\begin{figure}[hbtp]
\begin{center}
\mbox{\epsfxsize=144mm\epsffile{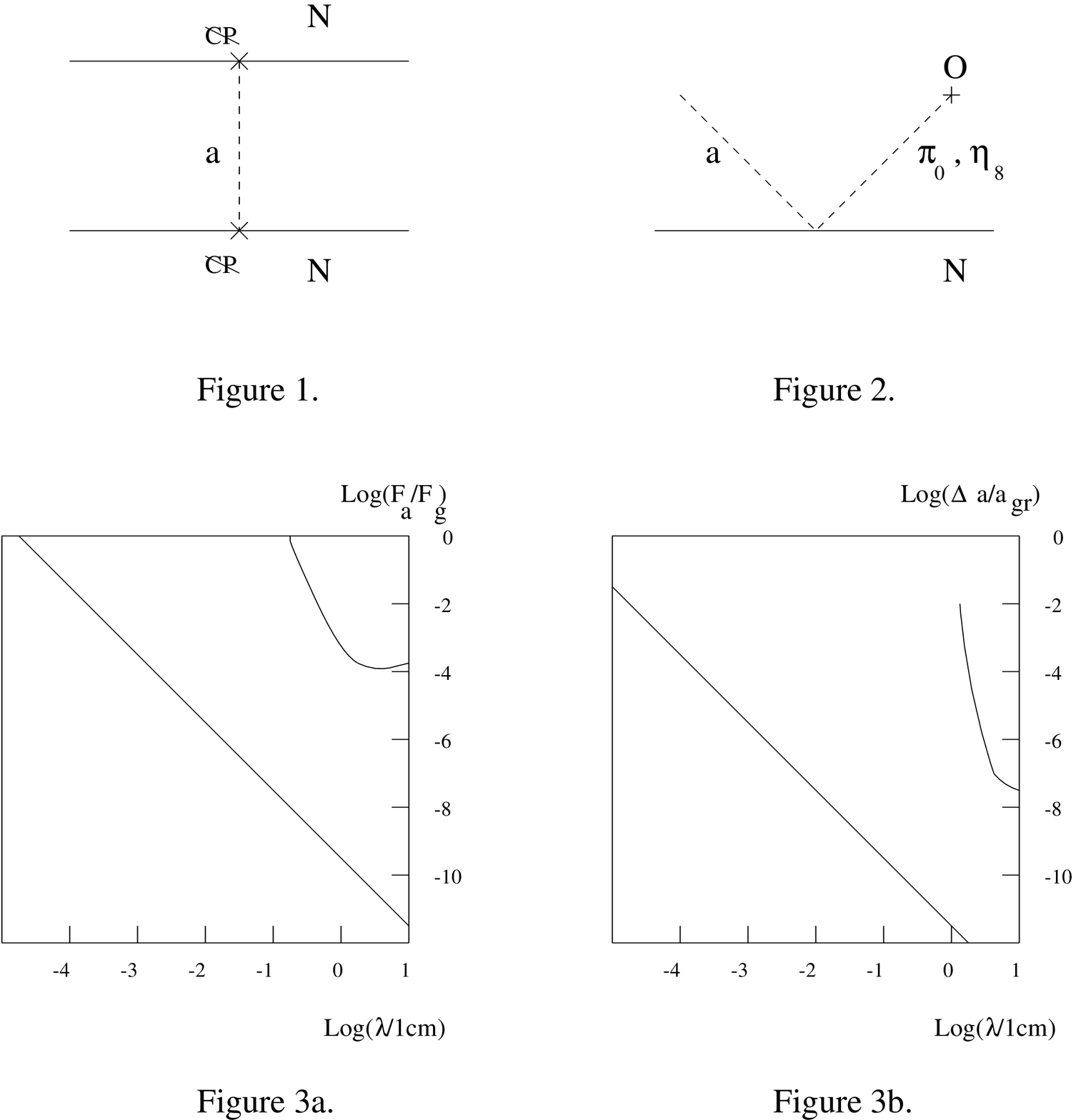}}
\end{center}
\label{muegf}
\end{figure}
\end{document}